\documentclass[12pt,a4paper]{article}

\usepackage{ifthen} 
\newboolean{pdflatex}
\setboolean{pdflatex}{true} 

\newboolean{articletitles}
\setboolean{articletitles}{true} 

\newboolean{uprightparticles}
\setboolean{uprightparticles}{false} 

\usepackage[top=1in, bottom=1.25in, left=1in, right=1in]{geometry}

\usepackage{microtype}
\usepackage{lineno}  
\usepackage{xspace} 
\usepackage{caption} 

\usepackage{graphicx}  
\usepackage{color}
\usepackage{colortbl}
\graphicspath{{./figs/}} 

\usepackage{amsmath} 
\usepackage{amssymb}
\usepackage{amsfonts}
\usepackage{upgreek} 

\newcommand*\patchAmsMathEnvironmentForLineno[1]{%
\expandafter\let\csname old#1\expandafter\endcsname\csname #1\endcsname
\expandafter\let\csname oldend#1\expandafter\endcsname\csname
end#1\endcsname
 \renewenvironment{#1}%
   {\linenomath\csname old#1\endcsname}%
   {\csname oldend#1\endcsname\endlinenomath}%
}
\newcommand*\patchBothAmsMathEnvironmentsForLineno[1]{%
  \patchAmsMathEnvironmentForLineno{#1}%
  \patchAmsMathEnvironmentForLineno{#1*}%
}
\AtBeginDocument{%
\patchBothAmsMathEnvironmentsForLineno{equation}%
\patchBothAmsMathEnvironmentsForLineno{align}%
\patchBothAmsMathEnvironmentsForLineno{flalign}%
\patchBothAmsMathEnvironmentsForLineno{alignat}%
\patchBothAmsMathEnvironmentsForLineno{gather}%
\patchBothAmsMathEnvironmentsForLineno{multline}%
\patchBothAmsMathEnvironmentsForLineno{eqnarray}%
}

\usepackage{hyperxmp}

\usepackage[pdftex]{hyperref}

\usepackage[colorinlistoftodos,textsize=scriptsize]{todonotes}

\usepackage[bottom,flushmargin,hang,multiple]{footmisc}

\usepackage[all]{hypcap} 

\usepackage{multirow}

\usepackage{xspace} 
\usepackage{upgreek}

\def\MagUp {\mbox{\em Mag\kern -0.05em Up}\xspace}

\ifthenelse{\boolean{uprightparticles}}%
{

 \def\PDelta      {\ensuremath{\Delta}\xspace}                 
 \def\PXi         {\ensuremath{\Xi}\xspace}                 
 \def\PLambda     {\ensuremath{\Lambda}\xspace}                 
 \def\PSigma      {\ensuremath{\Sigma}\xspace}                 
 \def\POmega      {\ensuremath{\Omega}\xspace}                 
 \def\PUpsilon    {\ensuremath{\Upsilon}\xspace}

 \def\PB      {\ensuremath{\mathrm{B}}\xspace}                 
                  
 \def\PD      {\ensuremath{\mathrm{D}}\xspace}

 \def\PK      {\ensuremath{\mathrm{K}}\xspace}

 \def\Pi      {\ensuremath{\mathrm{i}}\xspace}

 \def\Ps      {\ensuremath{\mathrm{s}}\xspace}

 \def\thebaroffset{0.0em}
}
{

 \mathchardef\PDelta="7101
 \mathchardef\PXi="7104
 \mathchardef\PLambda="7103
 \mathchardef\PSigma="7106
 \mathchardef\POmega="710A
 \mathchardef\PUpsilon="7107
                  
 \def\PB      {\ensuremath{B}\xspace}                 
                  
 \def\PD      {\ensuremath{D}\xspace}

 \def\PK      {\ensuremath{K}\xspace}

 \def\Pi      {\ensuremath{i}\xspace}

 \def\Ps      {\ensuremath{s}\xspace}

 \def\thebaroffset{0.18em}
}
\newcommand{\offsetoverline}[2][\thebaroffset]{\kern #1\overline{\kern -#1 #2}}%

\makeatletter
\ifcase \@ptsize \relax
  \newcommand{\miniscule}{\@setfontsize\miniscule{4}{5}}
\or
  \newcommand{\miniscule}{\@setfontsize\miniscule{5}{6}}
\or
  \newcommand{\miniscule}{\@setfontsize\miniscule{5}{6}}
\fi
\makeatother

\DeclareRobustCommand{\optbar}[1]{\shortstack{{\miniscule (\rule[.5ex]{1.25em}{.18mm})}
  \\ [-.7ex] $#1$}}

\def\squark    {{\ensuremath{\Ps}}\xspace}

\def\KorKbar {\kern \thebaroffset\optbar{\kern -\thebaroffset \PK}{}\xspace}

\def\D       {{\ensuremath{\PD}}\xspace}

\def\DorDbar {\kern \thebaroffset\optbar{\kern -\thebaroffset \PD}\xspace}

\def\Dp      {{\ensuremath{\D^+}}\xspace}
\def\Dm      {{\ensuremath{\D^-}}\xspace}

\def\DpDm    {\ensuremath{\Dp {\kern -0.16em \Dm}}\xspace}

\def\B       {{\ensuremath{\PB}}\xspace}

\def\BorBbar {\kern \thebaroffset\optbar{\kern -\thebaroffset \PB}\xspace}

\def\Bd      {{\ensuremath{\B^0}}\xspace}

\def\BdorBdbar {\kern \thebaroffset\optbar{\kern -\thebaroffset \Bd}\xspace}

\def\Bs      {{\ensuremath{\B^0_\squark}}\xspace}

\def\BsorBsbar {\kern \thebaroffset\optbar{\kern -\thebaroffset \Bs}\xspace}

\def\Y#1S{\ensuremath{\PUpsilon{(#1S)}}\xspace}

\def\LorLbar     {\kern \thebaroffset\optbar{\kern -\thebaroffset \PLambda}\xspace}

\def\to                 {\ensuremath{\rightarrow}\xspace}

\def\AT#1     {\ensuremath{A_{\mathrm{T}}^{#1}}\xspace}           

\def\C#1      {\ensuremath{\mathcal{C}_{#1}}\xspace}                       
\def\Cp#1     {\ensuremath{\mathcal{C}_{#1}^{'}}\xspace}                    
\def\Ceff#1   {\ensuremath{\mathcal{C}_{#1}^{\mathrm{(eff)}}}\xspace}        
\def\Cpeff#1  {\ensuremath{\mathcal{C}_{#1}^{'\mathrm{(eff)}}}\xspace}       
\def\Ope#1    {\ensuremath{\mathcal{O}_{#1}}\xspace}                       
\def\Opep#1   {\ensuremath{\mathcal{O}_{#1}^{'}}\xspace}                    

\newcommand{\aunit}[1]{\ensuremath{\text{\,#1}}}       

\newcommand{\tev}{\aunit{Te\kern -0.1em V}\xspace}
\newcommand{\gev}{\aunit{Ge\kern -0.1em V}\xspace}
\newcommand{\mev}{\aunit{Me\kern -0.1em V}\xspace}
\newcommand{\kev}{\aunit{ke\kern -0.1em V}\xspace}
\newcommand{\ev}{\aunit{e\kern -0.1em V}\xspace}
 
\newcommand{\mevc}{\ensuremath{\aunit{Me\kern -0.1em V\!/}c}\xspace}
\newcommand{\gevc}{\ensuremath{\aunit{Ge\kern -0.1em V\!/}c}\xspace}
\newcommand{\kevcc}{\ensuremath{\aunit{ke\kern -0.1em V\!/}c^2}\xspace}
\newcommand{\mevcc}{\ensuremath{\aunit{Me\kern -0.1em V\!/}c^2}\xspace}
\newcommand{\gevcc}{\ensuremath{\aunit{Ge\kern -0.1em V\!/}c^2}\xspace}

\def\gsim{{~\raise.15em\hbox{$>$}\kern-.85em
          \lower.35em\hbox{$\sim$}~}\xspace}
\def\lsim{{~\raise.15em\hbox{$<$}\kern-.85em
          \lower.35em\hbox{$\sim$}~}\xspace}

\def\tell1  {TELL1\xspace}
\def\ukl1   {UKL1\xspace}

\usepackage{cite} 
\usepackage{mciteplus}

\usepackage{graphicx}

\begin{document}

\begin{titlepage}

\vspace*{4cm}
{\huge\bf 
\begin{center}  
Sensitivity studies on the CKM angle $\gamma$ in $\Lambda_b^0 \to D\Lambda$ decays
\end{center}
}
\vspace{1.0cm}
\begin{center}
Shunan Zhang$^1$\footnote[1]{shunan@pku.edu.cn}, Yi Jiang$^2$\footnote[2]{jiangyi15@mails.ucas.ac.cn}, Zewen Chen$^2$\footnote[3]{chenzewen16@mails.ucas.ac.cn}, Wenbin Qian$^2$\footnote[4]{wenbin.qian@ucas.ac.cn} \\
\vspace{0.5cm}
{
\normalfont\itshape\footnotesize
$ ^1$State Key Laboratory of Nuclear Physics and Technology {\rm\&} School of Physics, Peking University, Beijing 100871, China\\
$ ^2$School of Physical Sciences, University of Chinese Academy of Sciences, Beijing 100049, China\\

}
\end{center}
\vspace{\fill}

\begin{abstract}
The sensitivity of the CKM angle $\gamma$ in $\Lambda_b^0 \to D \Lambda$ decays has been studied using the decay parameter $\alpha$ as an observable in addition to the decay rate asymmetry. Feasibility studies show that adding this observable improves the sensitivity on $\gamma$ by up to 60\% and makes the decays one of the most promising places to measure angle $\gamma$ and to search for $C\!P$ violation in $b$-baryon decays. 

\end{abstract}

\vspace{\fill}
\end{titlepage}


\setcounter{secnumdepth}{3}
\setcounter{tocdepth}{2}
\cleardoublepage

One of the primary goals of flavor physics is to search for extra sources of $C\!P$ violation by over-constraining the CKM matrix~\cite{PhysRevLett.10.531, Kobayashi:1973fv}.
The CKM angle, $\gamma \equiv \arg[-V_{ud}V_{ub}^*/V_{cd}V_{cb}^*]$, is one of the least known CKM parameters and limits the power of the global fit~\cite{CKMfitter2015,UTfit-UT}.
Many methods have been proposed to measure the angle $\gamma$ using tree-level transitions of $b \to c$ and $b \to u$ with negligible theoretical uncertainties~\cite{Brod:2013sga}. They are categorized according to the decays of $D$ meson. Throughout the paper, $D$ is used to indicate a quantum mixture of $D^0$ and $\overline{D}^0$. The GLW or ADS method~\cite{Gronau:1991dp,Gronau:1990ra,Atwood:1996ci} refers to $D$ decays into $C\!P$ eigenstates or non-charge-conjugated final states, where the favored (suppressed) $b$-decay is followed by a suppressed (favored) $D$ decay. The BPGGSZ method is used in case of multi-body $D$ decays, i.e. $K_{\mathrm{S}}^0\pi^+\pi^-$ or $K_{\mathrm{S}}^0K^+K^-$~\cite{Giri:2003ty,Bondar:2005ki,Bondar:2008hh}. 
The current world average for $\gamma$ from direct measurements is $(66.2^{+3.4}_{-3.6})^{\circ}$, dominated by the results from $B^+$ decays~\cite{HFLAV18}. Meanwhile the $\gamma$ values determined from global fits are $(65.6^{+0.9}_{-2.7})^\circ$ from CKMfitter~\cite{CKMfitter2015} and $(65.8\pm 2.2)^{\circ}$ from UTfit~\cite{UTfit-UT}.
Further improvement on direct measurements is eagerly desired. In addition, no $\gamma$ measurement has ever been performed in $b$-baryon decays. New measurements will not only add further precision on the global $\gamma$ determination from tree-level processes, but also help check if the CKM mechanism works in the baryon sector by comparing the $\gamma$ determined from the decays of $b$-baryons to that from $b$-mesons. 

Actually, $C\!P$ violation has only been found in meson decays, many searches in hyperon, $c$- or $b$-baryon decays have been extensively performed~\cite{LHCb-PAPER-2020-017,LHCb-PAPER-2019-028, LHCb-PAPER-2018-044,LHCb-PAPER-2018-025, LHCb-PAPER-2018-001, LHCb-PAPER-2019-026,LHCb-PAPER-2017-044, BESIII:2018cnd, BESIII:2020fqg, HyperCP:2004zvh} in the past ten years, no $C\!P$ violation is observed yet. Finding $C\!P$ violation in baryon decays will be an important step towards understanding the large matter-antimatter asymmetry observed in Universe~\cite{Gavela:1993ts}. The current focus on search for $C\!P$ violation in $b$-baryon decays are in charmless final states, however, when the $\gamma$ determined in $b$-baryon decays into charmed particles is statistically inconsistent with zero, it will also be an unambiguous sign of $C\!P$ violation.
The ADS decay of $\Lambda_b^0 \to D(K^+\pi^-)pK^-$ has recently been observed and $C\!P$ violation parameters are measured~\cite{LHCb-PAPER-2021-027}. Throughout the paper, charge-conjugation is applied if not specified. Extracting the angle $\gamma$ from $\Lambda_b^0 \to DpK^-$ decays are complicated due to many resonances involved. The spin-1/2 nature of $\Lambda_b^0$ and $p$ baryons makes it even more difficult than the partner decays of $B^+ \to D K^+$. 
Additional strong parameters have to be added to take into account different angular momentum contributions to the decay amplitudes and thus dilute the sensitivity to $\gamma$.

In this paper, we perform the sensitivity studies on the angle $\gamma$ in $\Lambda_b^0 \to D \Lambda$ decays, which have similar Feynman diagrams as $\Lambda_b^0 \to DpK^-$ in the $\Lambda^*$ region. Unlike $\Lambda_b^0 \to DpK^-$, 
an additional observable can be measured adding further constraints to $\gamma$~\cite{Giri:2001ju}. 
 
The angular momentum between $D$ and $\Lambda$ could be either $S$-wave or $P$-wave and one has to take both into account when measuring the angle $\gamma$.
The $S$-wave decay amplitudes are written as 
\begin{eqnarray}
    A_S(\Lambda_b^0 \to D^0 \Lambda) &=& A_{c,S}, \\
    A_S(\Lambda_b^0 \to \overline{D}^0 \Lambda) &=& A_{c,S}r_{B,S}e^{i (\delta_{B,S} -\gamma)},
\end{eqnarray}
for the $b\to c$ and $b\to u$ processes, respectively. The parameters $r_{B,S}$ and $\delta_{B,S}$ describe the strong interaction parts of the amplitude ratio between the two processes.
The corresponding $P$-wave amplitudes are denoted by replacing the subscript of $S$ into $P$. For the charge-conjugated decay amplitudes, the weak phase $\gamma$ changes the sign while other parts remain the same. 

The $D^0$ and $\overline{D}^0$ decay amplitudes into a final state $f$ are given by $A_D(f)$ and $\bar{A}_D(f)$, respectively. Studies in $B^+ \to D K^+$ have shown that bias on $\gamma$ caused by ignoring  $D^0$-$\overline{D}^0$ mixing effect and $C\!P$ violation in $D$ decays is less than 1$^{\circ}$~\cite{Grossman:2005rp,Rama:2013voa,Wang:2012ie}. In the recent LHCb $\gamma$ combination~\cite{LHCb-PAPER-2021-033}, they find that by fitting the $\gamma$ and $D$ mixing and $C\!P$ violation observables together, additional constraints on the $D$ mixing parameters are obtained. In this study, we neglect these contributions since their effects on $\gamma$ are marginal for the current small statistics of $\Lambda_b^0 \to D \Lambda$ decays and focus more on methodology itself. In the assumption, we have $\bar{A}_D(\bar{f}) = A_D(f)$ and $\bar{A}_D(f) = A_D(\bar{f})$. The full decay amplitude of $S$-wave is 
\begin{equation}
  A_S = A_{c,S} [A_D(f) + r_{B,S} e^{i (\delta_{B,S} - \gamma)} \bar{A}_D(f)], \end{equation}
and the decay rate of $\Lambda_b^0 \to D(f)\Lambda$ is given by
\begin{eqnarray}\label{Eq:decayrate}
    \Gamma(\Lambda_b^0 \to D(f)\Lambda) &=&|A_S|^2 + |A_P|^2 \\\nonumber
    &\propto& 
    (1+r^2)|A_D(f)|^2 + (r_{B,S}^2+r^2r_{B,P}^2) |\bar{A}_D(f)|^2 \\\nonumber
    &+& 2[r_{B,S}\Re(e^{i(\delta_{B,S}-\gamma)}\bar{A}_D A_D^*)+r_{B,P}r^2\Re(e^{i(\delta_{B,P}-\gamma)}\bar{A}_D A_D^*)].
\end{eqnarray}
The factor $|A_{c,S}|^2$ is omitted for simplicity and $A_{c,P}$ is replaced by $re^{i\delta} A_{c,S}$. The parameters $r$ and $\delta$ are to be determined from data.
Comparing Eq.~\ref{Eq:decayrate} with that in $B^+ \to DK^+$ decays~\cite{LHCb-PAPER-2021-033}, one can immediately find the decay rate in $\Lambda_b^0$ decays are more complicated and three additional strong parameters, $r$, $r_{B,P}$ and $\delta_{B,P}$, have to be introduced to take into account contributions from $P$-wave. The subsequent decay of $\Lambda \to p\pi^-$ only contributes an overall factor of $|A_S^{\Lambda}|^2 + |A_P^{\Lambda}|^2$ to the decay rate, where $A_S^{\Lambda}$ and $A_P^{\Lambda}$ are the $S$- and $P$-wave amplitudes of $\Lambda \to p\pi^-$, respectively, and are not listed.

There are other observables proposed to study $C\!P$ violation in hyperon decays~\cite{Lee:1957he,Lee:1957qs}, the decay parameters. They have been proposed previously to measure the angle $\gamma$ in $\Lambda_b^0 \to D\Lambda$ decays~\cite{Giri:2001ju}.
The decay parameters contain interference information between $S$- and $P$-waves. 
As $\Lambda_b^0$ produced in $pp$ collisions at LHC is found to be unpolarized~\cite{LHCb-PAPER-2020-005}, the only relevant decay parameter for $\Lambda_b^0 \to D\Lambda$ is 
\begin{eqnarray}
\alpha &=& \frac{2\Re(A_S^*A_P)}{|A_S|^2 + |A_P|^2}.
\end{eqnarray}
In the paper $\alpha_{-(+)}$ is used to denote $\alpha$ obtained from $\Lambda_b^0$ ($\overline{\Lambda}_b^0$) decays. 
The other two decay parameters $\beta = 2\Im(A_S^*A_P)/(|A_S|^2 + |A_P|^2)$ and $\gamma'= (|A_S|^2 - |A_P|^2)/(|A_S|^2 + |A_P|^2)$ can be measured if $\Lambda_b^0$ was polarized, which can also add constraints to the measurements of the angle $\gamma$~\cite{Giri:2001ju}. We use $\gamma'$ for the decay parameter to distinguish it from the angle $\gamma$.  

The decay parameter $\alpha$ can be measured using angular distributions 
\begin{equation}
    \frac{\textrm{d}\Gamma(\Lambda_b^0 \to D\Lambda(p\pi^-))}{\textrm{d}\Phi} \propto 1+ \alpha_-^{\Lambda_b} \alpha_-^{\Lambda} \cos \theta.
\end{equation}
Here $\theta$ is the helicity angle of the $\Lambda \to p\pi^-$ decay, defined as the direction of $\pi^-$ with respect to the direction of $D$ in the rest frame of $\Lambda$. 
The $\alpha_-^{\Lambda_b}$ and $\alpha_-^{\Lambda}$ are the decay parameters of the $\Lambda_b^0$ and $\Lambda$ decays. The $\alpha_-^{\Lambda}$ has been measured very precisely~\cite{PDG2020} and $\alpha_-^{\Lambda_b}$ is determined uniquely.
This offers us additional observables to constrain the angle $\gamma$ as
\begin{eqnarray}\label{Eq:interf}
    \Re(A_S^*A_P) &\propto& r [ |A_D|^2 \cos\delta + r_{B,S}r_{B,P}\cos(\delta+\delta_{B,P}-\delta_{B,S})|\bar{A}_D|^2 \\\nonumber
    & +& \Re(  r_{B,S}e^{-i(\delta_{B,S}-\gamma-\delta)}A_D\bar{A}_D^* + r_{B,P}e^{i(\delta_{B,P}-\gamma+\delta)}A_D^*\bar{A}_D )]. 
\end{eqnarray}
Combing Eq.~\ref{Eq:interf} and Eq.~\ref{Eq:decayrate}, the formula for $\alpha_{-}^{\Lambda_b}$ is obtained. In the above formula, 
despite the additional parameter $\delta$, there are more observables when considering the many decays in this study and their charge-conjugated ones. When $\alpha$ in different $D$ decays are the same, one obtains $r_{B,S}=r_{B,P}$ and $\delta_{B,S} = \delta_{B,P}$, the complications of $\Lambda_b^0$ decays degenerate and the sensitivity on $\gamma$ from $\Lambda_b^0 \to D \Lambda$ can be considered in a similar way as in $B^+ \to DK^+$. However, $r_B$ ($\sim 0.4$) in $\Lambda_b^0 \to DpK$ is larger than that in $B^+$ ($\sim 0.1$) ~\cite{LHCb-PAPER-2021-033} and results in larger $C\!P$ violation effects.

Sensitivities on $\gamma$ are studied using the decay rates, with and without the decay parameter $\alpha$. The decay channels of $D$ mesons considered are $D \to K^+K^-$, $D\to\pi^+ \pi^-$, $D \to K^{\pm}\pi^{\mp}$, $D \to K_{\mathrm{S}}^0 \pi^+\pi^-$ and $D \to K^{\pm}\pi^{\mp}\pi^+\pi^-$.  
For multi-body final states, 
amplitude variations over the phase space needs to be taken into account. It could be done either by using an amplitude model to describe the $D$ decays or by a model-independent method where the $D$ decay information is obtained from charm factories using quantum correlated $D$ samples from $\psi(3770) \to D^0 \overline{D}^0$. The model-independent method has the advantage of small systematic uncertainties and is more preferred for precise measurements. In this paper, we take the model-independent strategy where the phase space of $D$ decays is divided into different bins and only the integrated information in these bins 
\begin{equation}
   c_i + i s_i \equiv R_D^{i} e^{i\delta_D^{i}}  \equiv \frac{\int_{\textrm{bin i}} A_D(f) \overline{A}^*_D(f) \textrm{d}\Phi}{\sqrt{\int_{\textrm{bin i}} |A_D(f)|^2 \textrm{d}\Phi} \sqrt{\int_{\textrm{bin i}} |\overline{A}_D(f)|^2\textrm{d}\Phi}}
\end{equation}
are used.
The strong phase parameters $c_i$ and $s_i$ of $D \to K_{\mathrm{S}}^{0}\pi^+\pi^-$ decays are measured by CLEO-c and BESIII experiments ~\cite{CLEO:2010iul, BESIII:2020hlg, BESIII:2020khq} and the coherent factors $R_D$ and effective strong phase difference $\delta_D^i$ for $D \to K^{\mp}\pi^{\pm}\pi^+\pi^-$ are measured using the CLEO-c and BESIII data~\cite{Evans:2019wza,BESIII:2021eud}. The combinations of the results from the two experiments have also been performed~\cite{BESIII:2020hlg, BESIII:2020khq,BESIII:2021eud}.

The $\Lambda_b^0 \to D\Lambda$ has not been observed yet. Predictions based on factorization approach~\cite{Giri:2001ju,Hsiao:2015cda, Wang:2008sm, Wang:2015ndk} indicates that the branching fraction of $\Lambda_b^0 \to D \Lambda$ is about 200 times smaller than that of $\Lambda_b^0 \to J/\psi \Lambda$. 
Around 300 $\Lambda_b^0 \to D(K^-\pi^+)\Lambda$ events are expected to be reconstructed based on the data collected by LHCb in the years 2011-2018~\cite{LHCb-PAPER-2020-005}, and  is used for the sensitivity estimation.  
 The numbers for the other channels are scaled accordingly based on the formalism described previously. However, the yields of $D \to K_{\mathrm{S}}^{0} \pi^+\pi^-$ and $D \to K^{\mp}\pi^{\pm}\pi^+\pi^-$ are further scaled by a factor of 0.16 and 0.18~\cite{LHCb-PAPER-2020-036, LHCb-PAPER-2020-019, LHCb-PAPER-2016-003}  to take into account their efficiency differences with respect to the two-body decays.

In the study, $r_{B,S}$ and $r_{B,P}$ are set to 0.4 and 0.3, respectively, as both $\Lambda_b^0 \to D^0 \Lambda$ and $\Lambda_b^0 \to \overline{D}^0 \Lambda$ decays are color suppressed. The $r$ value describes the amplitude ratio between $S$- and $P$-waves and three sets of values (0.5, 1 or 2) are chosen. The three phases, $\delta$, $\delta_{B,S}$ and $\delta_{B,P}$, are also set to three different values, $0^{\circ}$, $60^{\circ}$ and $150^{\circ}$. 
The $\gamma$ input value is chosen to be $65^{\circ}$ and the charm inputs are taken from their averaged values ~\cite{PDG2020} or from those measured with CLEO-c or BESIII data~\cite{CLEO:2010iul, Evans:2019wza, BESIII:2020hlg, BESIII:2020khq, BESIII:2021eud}.

Pseudo data samples are generated according to the yields and $\alpha$ obtained from a set of parameters. A $\chi^2$ fit is then performed, where yields and $\alpha$ are considered to be independent as is shown in toys. In addition to the physical parameters, normalization factors are also considered as free parameters. The two-body $D$ decays share the same normalization factor following what has been done in real measurements, while the $K_{\mathrm{S}}^0\pi^+\pi^-$ and $K^{\mp}\pi^{\pm}\pi^+\pi^-$ have standalone normalization factors. This procedure is performed 1000 times for each set of inputs, and the uncertainty of $\gamma$ is obtained by a Gaussian fit to its distribution from the toy samples. No bias is found in all the parameter settings.
The procedures have been validated using parameters found in $B^+ \to DK^+$ decays~\cite{LHCb-PAPER-2021-033} with $r=0$. An uncertainty of 2.4$^{\circ}$ for the angle $\gamma$ is achieved. The slightly better precision is due to ignoring background contributions.

\begin{figure}[tb]
    \centering
    \includegraphics[width=0.49\textwidth]{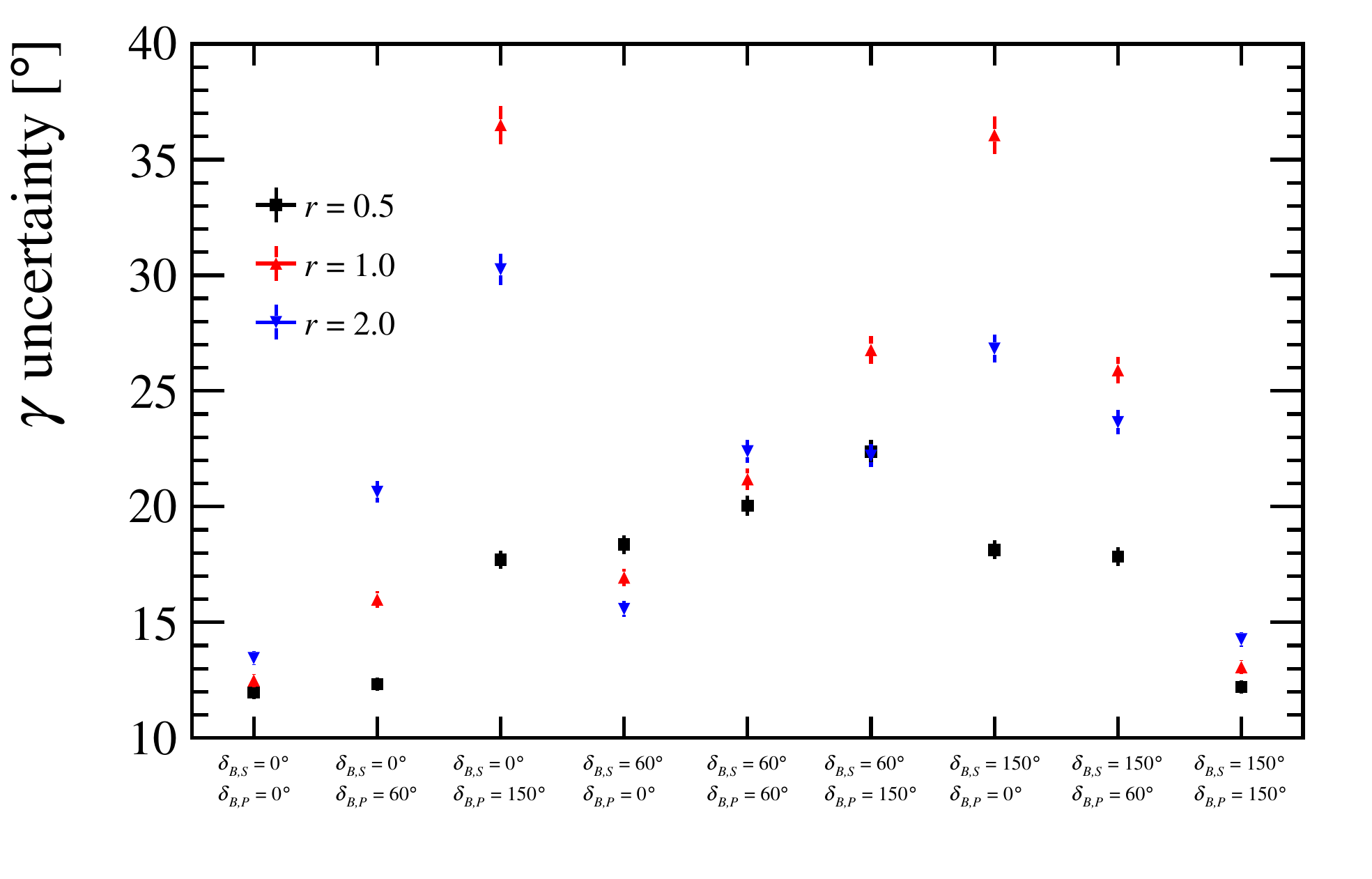}
    \includegraphics[width=0.49\textwidth]{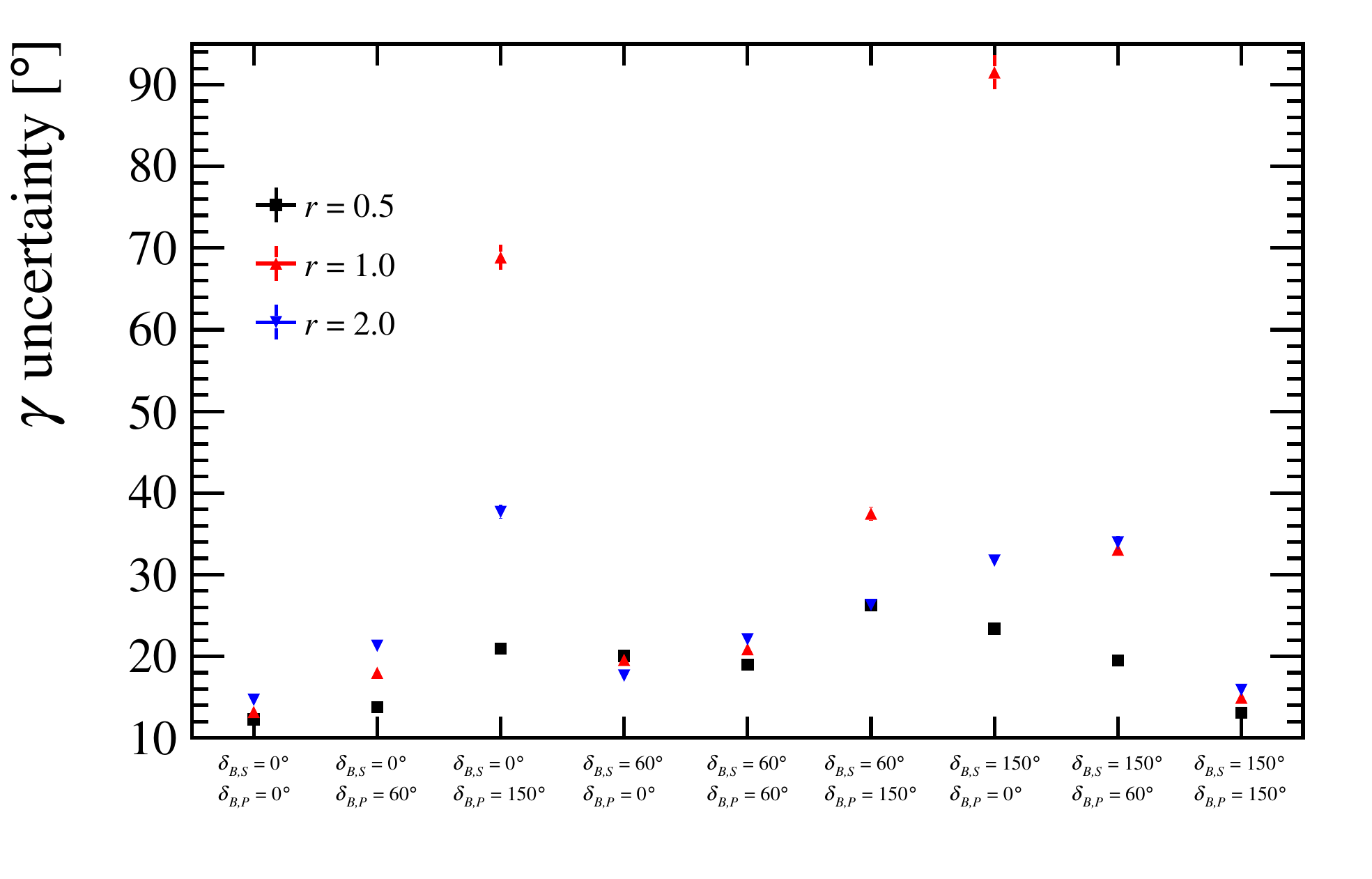}
    \caption{Statistical uncertainties of $\gamma$ from feasibility studies (left) with and (right) without the decay parameter $\alpha$. The parameters $r_{B,S}$,  $r_{B, P}$  and $\delta$  are set to be 0.4, 0.3 and 60$^{\circ}$, respectively, for all toy samples. }
    \label{fig:toyresult}
\end{figure}

Based on the above settings, the sensitivity of $\gamma$ in $\Lambda_b^0 \to D \Lambda$ decays is found to be around $12-36^{\circ}$ as shown in Fig.~\ref{fig:toyresult}.  
Dependence on the input values of $r, \delta_{B,S}$ and $\delta_{B,P}$ are found, while little on $\delta$. 
The sensitivity on $\gamma$ is improved by up to 60\% when including the observable $\alpha$. Using the current data collected by the LHCb experiment, it may be possible to observe $C\!P$ violation in $\Lambda_b^0 \to D\Lambda$ decays, and a sensitivity around $4-11^{\circ}$ will be achievable  with the data collected in the LHCb upgrade I period~\cite{LHCb-TDR-012}.
We strongly suggest the LHCb experiment performing the measurements proposed in this paper. 

~\\

\textbf{Note added:} 

After the first submission to arXiv, we are informed by Prof. Guy Wilkinson and Anton Poluektov that the idea to use the decay parameters in $\Lambda_b^0 \to D\Lambda$ decays has been proposed previously in hep-ph/0112220 and published in Phys. Rev. D65 (2002) 073029. The paper described the possibility of extracting the weak phase $\gamma$ using the decay parameters $\alpha'$, $\beta'$ and $\gamma'$ together with the decay rate. 
In the updated version of our paper, we modify our text (and title) to reflect this fact. 

In addition to the proposal of using the decay parametes as additional observables for $\gamma$ measurement,  we point out that only $\alpha$ can be measured as $\Lambda_b^0$ from $pp$ collisions are unpolarized and explain the experimental method to extract $\alpha$.  We perform the first sensitivity study with and without the decay parameter $\alpha$, proving that $\gamma$ can be extracted in the $\Lambda_b^0 \to D\Lambda$ decays. Including the observable $\alpha$ can improve the sensitivity by up to 60\%, depending on the strong parameters. The sensitivities for the current LHCb data and also the data collected during its upgrade are also given. In the sensitivity study, additional channels of $D \to K_S \pi^+ \pi^-$ and $D \to K^{\pm}\pi^{\mp} \pi^+ \pi^-$ are considered.

\section*{Acknowledgements}

This work is partially supported by 
National Natural Science Foundation of China (NSFC) under grant No.  11975015, 
the Fundamental Research Funds for the Central Universities.
We thank Prof. Haibo Li, Yuehong Xie and Yuming Wang for useful discussions on the topic. We thank Prof. Guy Wilkinson and Anton Poluektov for pointing out the previous work and for their insight suggestions to improve the manuscripts.

\bibliographystyle{LHCb}
\bibliography{main,standard,LHCb-PAPER,LHCb-CONF,LHCb-DP,LHCb-TDR}

\ifx\mcitethebibliography\mciteundefinedmacro
\PackageError{LHCb.bst}{mciteplus.sty has not been loaded}
{This bibstyle requires the use of the mciteplus package.}\fi
\providecommand{\href}[2]{#2}
\begin{mcitethebibliography}{10}
\mciteSetBstSublistMode{n}
\mciteSetBstMaxWidthForm{subitem}{\alph{mcitesubitemcount})}
\mciteSetBstSublistLabelBeginEnd{\mcitemaxwidthsubitemform\space}
{\relax}{\relax}

\bibitem{PhysRevLett.10.531}
N.~Cabibbo, \ifthenelse{\boolean{articletitles}}{\emph{Unitary symmetry and
  leptonic decays}, }{}\href{https://doi.org/10.1103/PhysRevLett.10.531}{Phys.\
  Rev.\ Lett.\  \textbf{10} (1963) 531}\relax
\mciteBstWouldAddEndPuncttrue
\mciteSetBstMidEndSepPunct{\mcitedefaultmidpunct}
{\mcitedefaultendpunct}{\mcitedefaultseppunct}\relax
\EndOfBibitem
\bibitem{Kobayashi:1973fv}
M.~Kobayashi and T.~Maskawa, \ifthenelse{\boolean{articletitles}}{\emph{{CP
  Violation in the Renormalizable Theory of Weak Interaction}},
  }{}\href{https://doi.org/10.1143/PTP.49.652}{Prog.\ Theor.\ Phys.\
  \textbf{49} (1973) 652}\relax
\mciteBstWouldAddEndPuncttrue
\mciteSetBstMidEndSepPunct{\mcitedefaultmidpunct}
{\mcitedefaultendpunct}{\mcitedefaultseppunct}\relax
\EndOfBibitem
\bibitem{CKMfitter2015}
CKMfitter group, J.~Charles {\em et~al.},
  \ifthenelse{\boolean{articletitles}}{\emph{{Current status of the standard
  model CKM fit and constraints on \hbox{$\Delta F=2$} new physics}},
  }{}\href{https://doi.org/10.1103/PhysRevD.91.073007}{Phys.\ Rev.\
  \textbf{D91} (2015) 073007},
  \href{http://arxiv.org/abs/1501.05013}{{\normalfont\ttfamily
  arXiv:1501.05013}}, {updated results and plots available at
  \href{http://ckmfitter.in2p3.fr/}{{\texttt{http://ckmfitter.in2p3.fr/}}}}\relax
\mciteBstWouldAddEndPuncttrue
\mciteSetBstMidEndSepPunct{\mcitedefaultmidpunct}
{\mcitedefaultendpunct}{\mcitedefaultseppunct}\relax
\EndOfBibitem
\bibitem{UTfit-UT}
UTfit collaboration, M.~Bona {\em et~al.},
  \ifthenelse{\boolean{articletitles}}{\emph{{The unitarity triangle fit in the
  standard model and hadronic parameters from lattice QCD: A reappraisal after
  the measurements of $\Delta m_{s}$ and $BR(B\to\tau\nu_{\tau})$}},
  }{}\href{https://doi.org/10.1088/1126-6708/2006/10/081}{JHEP \textbf{10}
  (2006) 081}, \href{http://arxiv.org/abs/hep-ph/0606167}{{\normalfont\ttfamily
  arXiv:hep-ph/0606167}}, {updated results and plots available at
  \href{http://www.utfit.org/}{{\texttt{http://www.utfit.org/}}}}\relax
\mciteBstWouldAddEndPuncttrue
\mciteSetBstMidEndSepPunct{\mcitedefaultmidpunct}
{\mcitedefaultendpunct}{\mcitedefaultseppunct}\relax
\EndOfBibitem
\bibitem{Brod:2013sga}
J.~Brod and J.~Zupan, \ifthenelse{\boolean{articletitles}}{\emph{{The ultimate
  theoretical error on $\gamma$ from $B \to DK$ decays}},
  }{}\href{https://doi.org/10.1007/JHEP01(2014)051}{JHEP \textbf{01} (2014)
  051}, \href{http://arxiv.org/abs/1308.5663}{{\normalfont\ttfamily
  arXiv:1308.5663}}\relax
\mciteBstWouldAddEndPuncttrue
\mciteSetBstMidEndSepPunct{\mcitedefaultmidpunct}
{\mcitedefaultendpunct}{\mcitedefaultseppunct}\relax
\EndOfBibitem
\bibitem{Gronau:1991dp}
M.~Gronau and D.~Wyler, \ifthenelse{\boolean{articletitles}}{\emph{{On
  determining a weak phase from CP asymmetries in charged $B$ decays}},
  }{}\href{https://doi.org/10.1016/0370-2693(91)90034-N}{Phys.\ Lett.\ B
  \textbf{265} (1991) 172}\relax
\mciteBstWouldAddEndPuncttrue
\mciteSetBstMidEndSepPunct{\mcitedefaultmidpunct}
{\mcitedefaultendpunct}{\mcitedefaultseppunct}\relax
\EndOfBibitem
\bibitem{Gronau:1990ra}
M.~Gronau and D.~London, \ifthenelse{\boolean{articletitles}}{\emph{{How to
  determine all the angles of the unitarity triangle from $B_d^0 \rightarrow D
  K_s$ and $B_s^0 \rightarrow D\phi$}},
  }{}\href{https://doi.org/10.1016/0370-2693(91)91756-L}{Phys.\ Lett.\ B
  \textbf{253} (1991) 483}\relax
\mciteBstWouldAddEndPuncttrue
\mciteSetBstMidEndSepPunct{\mcitedefaultmidpunct}
{\mcitedefaultendpunct}{\mcitedefaultseppunct}\relax
\EndOfBibitem
\bibitem{Atwood:1996ci}
D.~Atwood, I.~Dunietz, and A.~Soni,
  \ifthenelse{\boolean{articletitles}}{\emph{{Enhanced CP violation with $B\to
  K\Dz(\Dzb)$ modes and extraction of the Cabibbo-Kobayashi-Maskawa angle
  $\gamma$}}, }{}\href{https://doi.org/10.1103/PhysRevLett.78.3257}{Phys.\
  Rev.\ Lett.\  \textbf{78} (1997) 3257},
  \href{http://arxiv.org/abs/hep-ph/9612433}{{\normalfont\ttfamily
  arXiv:hep-ph/9612433}}\relax
\mciteBstWouldAddEndPuncttrue
\mciteSetBstMidEndSepPunct{\mcitedefaultmidpunct}
{\mcitedefaultendpunct}{\mcitedefaultseppunct}\relax
\EndOfBibitem
\bibitem{Giri:2003ty}
A.~Giri, Y.~Grossman, A.~Soffer, and J.~Zupan,
  \ifthenelse{\boolean{articletitles}}{\emph{{Determining $\gamma$ using
  $\Bpm\to D\Kpm$ with multibody $D$ decays}},
  }{}\href{https://doi.org/10.1103/PhysRevD.68.054018}{Phys.\ Rev.\ D
  \textbf{68} (2003) 054018},
  \href{http://arxiv.org/abs/hep-ph/0303187}{{\normalfont\ttfamily
  arXiv:hep-ph/0303187}}\relax
\mciteBstWouldAddEndPuncttrue
\mciteSetBstMidEndSepPunct{\mcitedefaultmidpunct}
{\mcitedefaultendpunct}{\mcitedefaultseppunct}\relax
\EndOfBibitem
\bibitem{Bondar:2005ki}
A.~Bondar and A.~Poluektov,
  \ifthenelse{\boolean{articletitles}}{\emph{{Feasibility study of
  model-independent approach to $\varphi_3$ measurement using Dalitz plot
  analysis}}, }{}\href{https://doi.org/10.1140/epjc/s2006-02590-x}{Eur.\ Phys.\
  J.\ C \textbf{47} (2006) 347},
  \href{http://arxiv.org/abs/hep-ph/0510246}{{\normalfont\ttfamily
  arXiv:hep-ph/0510246}}\relax
\mciteBstWouldAddEndPuncttrue
\mciteSetBstMidEndSepPunct{\mcitedefaultmidpunct}
{\mcitedefaultendpunct}{\mcitedefaultseppunct}\relax
\EndOfBibitem
\bibitem{Bondar:2008hh}
A.~Bondar and A.~Poluektov, \ifthenelse{\boolean{articletitles}}{\emph{{The use
  of quantum-correlated $\Dz$ decays for $\varphi_3$ measurement}},
  }{}\href{https://doi.org/10.1140/epjc/s10052-008-0600-z}{Eur.\ Phys.\ J.\ C
  \textbf{55} (2008) 51},
  \href{http://arxiv.org/abs/0801.0840}{{\normalfont\ttfamily
  arXiv:0801.0840}}\relax
\mciteBstWouldAddEndPuncttrue
\mciteSetBstMidEndSepPunct{\mcitedefaultmidpunct}
{\mcitedefaultendpunct}{\mcitedefaultseppunct}\relax
\EndOfBibitem
\bibitem{HFLAV18}
Heavy Flavor Averaging Group, Y.~Amhis {\em et~al.},
  \ifthenelse{\boolean{articletitles}}{\emph{{Averages of $b$-hadron,
  $c$-hadron, and $\tau$-lepton properties as of 2018}},
  }{}\href{https://doi.org/10.1140/epjc/s10052-020-8156-7}{Eur.\ Phys.\ J.\
  \textbf{C81} (2021) 226},
  \href{http://arxiv.org/abs/1909.12524}{{\normalfont\ttfamily
  arXiv:1909.12524}}, {updated results and plots available at
  \href{https://hflav.web.cern.ch}{{\texttt{https://hflav.web.cern.ch}}}}\relax
\mciteBstWouldAddEndPuncttrue
\mciteSetBstMidEndSepPunct{\mcitedefaultmidpunct}
{\mcitedefaultendpunct}{\mcitedefaultseppunct}\relax
\EndOfBibitem
\bibitem{LHCb-PAPER-2020-017}
LHCb collaboration, R.~Aaij {\em et~al.},
  \ifthenelse{\boolean{articletitles}}{\emph{{Search for $CP$ violation in
  $\Xibm \to p K^- K^- decays$}},
  }{}\href{https://doi.org/10.1103/PhysRevD.104.052010}{Phys.\ Rev.\
  \textbf{D104} (2021) 052010},
  \href{http://arxiv.org/abs/2104.15074}{{\normalfont\ttfamily
  arXiv:2104.15074}}\relax
\mciteBstWouldAddEndPuncttrue
\mciteSetBstMidEndSepPunct{\mcitedefaultmidpunct}
{\mcitedefaultendpunct}{\mcitedefaultseppunct}\relax
\EndOfBibitem
\bibitem{LHCb-PAPER-2019-028}
LHCb collaboration, R.~Aaij {\em et~al.},
  \ifthenelse{\boolean{articletitles}}{\emph{{Search for \CP violation and
  observation of $P$ violation in \decay{\Lb}{p\pim\pip\pim} decays}},
  }{}\href{https://doi.org/10.1103/PhysRevD.102.051101}{Phys.\ Rev.\
  \textbf{D102} (2020) 051101},
  \href{http://arxiv.org/abs/1912.10741}{{\normalfont\ttfamily
  arXiv:1912.10741}}\relax
\mciteBstWouldAddEndPuncttrue
\mciteSetBstMidEndSepPunct{\mcitedefaultmidpunct}
{\mcitedefaultendpunct}{\mcitedefaultseppunct}\relax
\EndOfBibitem
\bibitem{LHCb-PAPER-2018-044}
LHCb collaboration, R.~Aaij {\em et~al.},
  \ifthenelse{\boolean{articletitles}}{\emph{{Measurement of \CP asymmetries in
  charmless four-body \Lb and \Xibz decays}},
  }{}\href{https://doi.org/10.1140/epjc/s10052-019-7218-1}{Eur.\ Phys.\ J.\
  \textbf{C79} (2019) 745},
  \href{http://arxiv.org/abs/1903.06792}{{\normalfont\ttfamily
  arXiv:1903.06792}}\relax
\mciteBstWouldAddEndPuncttrue
\mciteSetBstMidEndSepPunct{\mcitedefaultmidpunct}
{\mcitedefaultendpunct}{\mcitedefaultseppunct}\relax
\EndOfBibitem
\bibitem{LHCb-PAPER-2018-025}
LHCb collaboration, R.~Aaij {\em et~al.},
  \ifthenelse{\boolean{articletitles}}{\emph{{Search for \CP violation in
  \mbox{\decay{\Lb}{p\Km} and \mbox{\decay{\Lb}{p\pim}}} decays}},
  }{}\href{https://doi.org/10.1016/j.physletb.2018.10.039}{Phys.\ Lett.\
  \textbf{B784} (2018) 101},
  \href{http://arxiv.org/abs/1807.06544}{{\normalfont\ttfamily
  arXiv:1807.06544}}\relax
\mciteBstWouldAddEndPuncttrue
\mciteSetBstMidEndSepPunct{\mcitedefaultmidpunct}
{\mcitedefaultendpunct}{\mcitedefaultseppunct}\relax
\EndOfBibitem
\bibitem{LHCb-PAPER-2018-001}
LHCb collaboration, R.~Aaij {\em et~al.},
  \ifthenelse{\boolean{articletitles}}{\emph{{Search for \CP violation using
  triple product asymmetries in \mbox{\decay{\Lb}{p\Km\pip\pim}},
  \mbox{\decay{\Lb}{p\Km\Kp\Km}}, and \mbox{\decay{\Xires^0_b}{p\Km\Km\pip}}
  decays}}, }{}\href{https://doi.org/10.1007/JHEP08(2018)039}{JHEP \textbf{08}
  (2018) 039}, \href{http://arxiv.org/abs/1805.03941}{{\normalfont\ttfamily
  arXiv:1805.03941}}\relax
\mciteBstWouldAddEndPuncttrue
\mciteSetBstMidEndSepPunct{\mcitedefaultmidpunct}
{\mcitedefaultendpunct}{\mcitedefaultseppunct}\relax
\EndOfBibitem
\bibitem{LHCb-PAPER-2019-026}
LHCb collaboration, R.~Aaij {\em et~al.},
  \ifthenelse{\boolean{articletitles}}{\emph{{Search for \CP violation in
  \mbox{\decay{\Xicp}{p\Km\pip}} decays with model-independent techniques}},
  }{}\href{https://doi.org/10.1140/epjc/s10052-020-8365-0}{Eur.\ Phys.\ J.\
  \textbf{C80} (2020) 986},
  \href{http://arxiv.org/abs/2006.03145}{{\normalfont\ttfamily
  arXiv:2006.03145}}\relax
\mciteBstWouldAddEndPuncttrue
\mciteSetBstMidEndSepPunct{\mcitedefaultmidpunct}
{\mcitedefaultendpunct}{\mcitedefaultseppunct}\relax
\EndOfBibitem
\bibitem{LHCb-PAPER-2017-044}
LHCb collaboration, R.~Aaij {\em et~al.},
  \ifthenelse{\boolean{articletitles}}{\emph{{Search for \CP violation in
  \mbox{\decay{\Lc}{p \Km \Kp}} and \mbox{\decay{\Lc}{p\pim\pip}} decays}},
  }{}\href{https://doi.org/10.1007/JHEP03(2018)182}{JHEP \textbf{03} (2018)
  182}, \href{http://arxiv.org/abs/1712.07051}{{\normalfont\ttfamily
  arXiv:1712.07051}}\relax
\mciteBstWouldAddEndPuncttrue
\mciteSetBstMidEndSepPunct{\mcitedefaultmidpunct}
{\mcitedefaultendpunct}{\mcitedefaultseppunct}\relax
\EndOfBibitem
\bibitem{BESIII:2018cnd}
BESIII collaboration, M.~Ablikim {\em et~al.},
  \ifthenelse{\boolean{articletitles}}{\emph{{Polarization and entanglement in
  baryon-antibaryon pair production in electron-positron annihilation}},
  }{}\href{https://doi.org/10.1038/s41567-019-0494-8}{Nature Phys.\
  \textbf{15} (2019) 631},
  \href{http://arxiv.org/abs/1808.08917}{{\normalfont\ttfamily
  arXiv:1808.08917}}\relax
\mciteBstWouldAddEndPuncttrue
\mciteSetBstMidEndSepPunct{\mcitedefaultmidpunct}
{\mcitedefaultendpunct}{\mcitedefaultseppunct}\relax
\EndOfBibitem
\bibitem{BESIII:2020fqg}
BESIII collaboration, M.~Ablikim {\em et~al.},
  \ifthenelse{\boolean{articletitles}}{\emph{{$\Sigma^{+}$ and $\bar{\Sigma}^-$
  polarization in the $J/\psi$ and $\psi(3686)$ decays}},
  }{}\href{https://doi.org/10.1103/PhysRevLett.125.052004}{Phys.\ Rev.\ Lett.\
  \textbf{125} (2020) 052004},
  \href{http://arxiv.org/abs/2004.07701}{{\normalfont\ttfamily
  arXiv:2004.07701}}\relax
\mciteBstWouldAddEndPuncttrue
\mciteSetBstMidEndSepPunct{\mcitedefaultmidpunct}
{\mcitedefaultendpunct}{\mcitedefaultseppunct}\relax
\EndOfBibitem
\bibitem{HyperCP:2004zvh}
HyperCP collaboration, T.~Holmstrom {\em et~al.},
  \ifthenelse{\boolean{articletitles}}{\emph{{Search for CP violation in
  charged-$\Xi$ and $\Lambda$ hyperon decays}},
  }{}\href{https://doi.org/10.1103/PhysRevLett.93.262001}{Phys.\ Rev.\ Lett.\
  \textbf{93} (2004) 262001},
  \href{http://arxiv.org/abs/hep-ex/0412038}{{\normalfont\ttfamily
  arXiv:hep-ex/0412038}}\relax
\mciteBstWouldAddEndPuncttrue
\mciteSetBstMidEndSepPunct{\mcitedefaultmidpunct}
{\mcitedefaultendpunct}{\mcitedefaultseppunct}\relax
\EndOfBibitem
\bibitem{Gavela:1993ts}
M.~B. Gavela, P.~Hernandez, J.~Orloff, and O.~P\`ene,
  \ifthenelse{\boolean{articletitles}}{\emph{{Standard model \CP violation and
  baryon asymmetry}}, }{}\href{https://doi.org/10.1142/S0217732394000629}{Mod.\
  Phys.\ Lett.\  \textbf{A9} (1994) 795},
  \href{http://arxiv.org/abs/hep-ph/9312215}{{\normalfont\ttfamily
  arXiv:hep-ph/9312215}}\relax
\mciteBstWouldAddEndPuncttrue
\mciteSetBstMidEndSepPunct{\mcitedefaultmidpunct}
{\mcitedefaultendpunct}{\mcitedefaultseppunct}\relax
\EndOfBibitem
\bibitem{LHCb-PAPER-2021-027}
LHCb collaboration, R.~Aaij {\em et~al.},
  \ifthenelse{\boolean{articletitles}}{\emph{{Observation of the suppressed
  $\Lb \to \D \Pp \Km$ decay with $\D \to \Kp \pim$ and measurement of its CP
  asymmetry}}, }{}\href{http://arxiv.org/abs/2109.02621}{{\normalfont\ttfamily
  arXiv:2109.02621}}, {submitted to PRD}\relax
\mciteBstWouldAddEndPuncttrue
\mciteSetBstMidEndSepPunct{\mcitedefaultmidpunct}
{\mcitedefaultendpunct}{\mcitedefaultseppunct}\relax
\EndOfBibitem
\bibitem{Giri:2001ju}
A.~K. Giri, R.~Mohanta, and M.~P. Khanna,
  \ifthenelse{\boolean{articletitles}}{\emph{{Possibility of extracting the
  weak phase $\gamma$ from $\Lambda_b \to \Lambda D^0$ decays}},
  }{}\href{https://doi.org/10.1103/PhysRevD.65.073029}{Phys.\ Rev.\ D
  \textbf{65} (2002) 073029},
  \href{http://arxiv.org/abs/hep-ph/0112220}{{\normalfont\ttfamily
  arXiv:hep-ph/0112220}}\relax
\mciteBstWouldAddEndPuncttrue
\mciteSetBstMidEndSepPunct{\mcitedefaultmidpunct}
{\mcitedefaultendpunct}{\mcitedefaultseppunct}\relax
\EndOfBibitem
\bibitem{Grossman:2005rp}
Y.~Grossman, A.~Soffer, and J.~Zupan,
  \ifthenelse{\boolean{articletitles}}{\emph{{The Effect of $D-\bar{D}$ mixing
  on the measurement of $\gamma$ in $B\to DK$ decays}},
  }{}\href{https://doi.org/10.1103/PhysRevD.72.031501}{Phys.\ Rev.\ D
  \textbf{72} (2005) 031501},
  \href{http://arxiv.org/abs/hep-ph/0505270}{{\normalfont\ttfamily
  arXiv:hep-ph/0505270}}\relax
\mciteBstWouldAddEndPuncttrue
\mciteSetBstMidEndSepPunct{\mcitedefaultmidpunct}
{\mcitedefaultendpunct}{\mcitedefaultseppunct}\relax
\EndOfBibitem
\bibitem{Rama:2013voa}
M.~Rama, \ifthenelse{\boolean{articletitles}}{\emph{{Effect of $D-\bar{D}$
  mixing in the extraction of $\gamma$ with $B^- \to D^0 K^-$ and $B^- \to D^0
  \pi^-$ decays}}, }{}\href{https://doi.org/10.1103/PhysRevD.89.014021}{Phys.\
  Rev.\ D \textbf{89} (2014) 014021},
  \href{http://arxiv.org/abs/1307.4384}{{\normalfont\ttfamily
  arXiv:1307.4384}}\relax
\mciteBstWouldAddEndPuncttrue
\mciteSetBstMidEndSepPunct{\mcitedefaultmidpunct}
{\mcitedefaultendpunct}{\mcitedefaultseppunct}\relax
\EndOfBibitem
\bibitem{Wang:2012ie}
W.~Wang, \ifthenelse{\boolean{articletitles}}{\emph{{CP violation effects on
  the measurement of the Cabibbo-Kobayashi-Maskawa angle $\gamma$ from $B \to D
  K$}}, }{}\href{https://doi.org/10.1103/PhysRevLett.110.061802}{Phys.\ Rev.\
  Lett.\  \textbf{110} (2013) 061802},
  \href{http://arxiv.org/abs/1211.4539}{{\normalfont\ttfamily
  arXiv:1211.4539}}\relax
\mciteBstWouldAddEndPuncttrue
\mciteSetBstMidEndSepPunct{\mcitedefaultmidpunct}
{\mcitedefaultendpunct}{\mcitedefaultseppunct}\relax
\EndOfBibitem
\bibitem{LHCb-PAPER-2021-033}
LHCb collaboration, R.~Aaij {\em et~al.},
  \ifthenelse{\boolean{articletitles}}{\emph{{First simultaneous determination
  of CKM angle $\gamma$ and charm mixing parameters from a combination of LHCb
  measurements}},
  }{}\href{http://arxiv.org/abs/2110.02350}{{\normalfont\ttfamily
  arXiv:2110.02350}}, {submitted to JHEP}\relax
\mciteBstWouldAddEndPuncttrue
\mciteSetBstMidEndSepPunct{\mcitedefaultmidpunct}
{\mcitedefaultendpunct}{\mcitedefaultseppunct}\relax
\EndOfBibitem
\bibitem{Lee:1957he}
T.~D. Lee {\em et~al.}, \ifthenelse{\boolean{articletitles}}{\emph{{Possible
  Detection of Parity Nonconservation in Hyperon Decay}},
  }{}\href{https://doi.org/10.1103/PhysRev.106.1367}{Phys.\ Rev.\  \textbf{106}
  (1957) 1367}\relax
\mciteBstWouldAddEndPuncttrue
\mciteSetBstMidEndSepPunct{\mcitedefaultmidpunct}
{\mcitedefaultendpunct}{\mcitedefaultseppunct}\relax
\EndOfBibitem
\bibitem{Lee:1957qs}
T.~D. Lee and C.-N. Yang, \ifthenelse{\boolean{articletitles}}{\emph{{General
  Partial Wave Analysis of the Decay of a Hyperon of Spin 1/2}},
  }{}\href{https://doi.org/10.1103/PhysRev.108.1645}{Phys.\ Rev.\  \textbf{108}
  (1957) 1645}\relax
\mciteBstWouldAddEndPuncttrue
\mciteSetBstMidEndSepPunct{\mcitedefaultmidpunct}
{\mcitedefaultendpunct}{\mcitedefaultseppunct}\relax
\EndOfBibitem
\bibitem{LHCb-PAPER-2020-005}
LHCb collaboration, R.~Aaij {\em et~al.},
  \ifthenelse{\boolean{articletitles}}{\emph{{Measurement of the $\Lb \to \jpsi
  \Lz$ angular distribution and the \Lz polarisation in $pp$ collisions}},
  }{}\href{https://doi.org/10.1007/JHEP06(2020)110}{JHEP \textbf{06} (2020)
  110}, \href{http://arxiv.org/abs/2004.10563}{{\normalfont\ttfamily
  arXiv:2004.10563}}\relax
\mciteBstWouldAddEndPuncttrue
\mciteSetBstMidEndSepPunct{\mcitedefaultmidpunct}
{\mcitedefaultendpunct}{\mcitedefaultseppunct}\relax
\EndOfBibitem
\bibitem{PDG2020}
Particle Data Group, P.~A. Zyla {\em et~al.},
  \ifthenelse{\boolean{articletitles}}{\emph{{\href{http://pdg.lbl.gov/}{Review
  of particle physics}}}, }{}\href{https://doi.org/10.1093/ptep/ptaa104}{Prog.\
  Theor.\ Exp.\ Phys.\  \textbf{2020} (2020) 083C01}\relax
\mciteBstWouldAddEndPuncttrue
\mciteSetBstMidEndSepPunct{\mcitedefaultmidpunct}
{\mcitedefaultendpunct}{\mcitedefaultseppunct}\relax
\EndOfBibitem
\bibitem{CLEO:2010iul}
CLEO collaboration, J.~Libby {\em et~al.},
  \ifthenelse{\boolean{articletitles}}{\emph{{Model-independent determination
  of the strong-phase difference between $D^0$ and $\bar{D}^0 \to K^0_{S,L} h^+
  h^-$ ($h=\pi,K$) and its impact on the measurement of the CKM angle
  $\gamma/\phi_3$}},
  }{}\href{https://doi.org/10.1103/PhysRevD.82.112006}{Phys.\ Rev.\ D
  \textbf{82} (2010) 112006},
  \href{http://arxiv.org/abs/1010.2817}{{\normalfont\ttfamily
  arXiv:1010.2817}}\relax
\mciteBstWouldAddEndPuncttrue
\mciteSetBstMidEndSepPunct{\mcitedefaultmidpunct}
{\mcitedefaultendpunct}{\mcitedefaultseppunct}\relax
\EndOfBibitem
\bibitem{BESIII:2020hlg}
BESIII collaboration, M.~Ablikim {\em et~al.},
  \ifthenelse{\boolean{articletitles}}{\emph{{Determination of Strong-Phase
  Parameters in $D\rightarrow K^0_{S,L}\pi^+\pi^-$}},
  }{}\href{https://doi.org/10.1103/PhysRevLett.124.241802}{Phys.\ Rev.\ Lett.\
  \textbf{124} (2020) 241802},
  \href{http://arxiv.org/abs/2002.12791}{{\normalfont\ttfamily
  arXiv:2002.12791}}\relax
\mciteBstWouldAddEndPuncttrue
\mciteSetBstMidEndSepPunct{\mcitedefaultmidpunct}
{\mcitedefaultendpunct}{\mcitedefaultseppunct}\relax
\EndOfBibitem
\bibitem{BESIII:2020khq}
BESIII collaboration, M.~Ablikim {\em et~al.},
  \ifthenelse{\boolean{articletitles}}{\emph{{Model-independent determination
  of the relative strong-phase difference between $D^0$ and
  $\bar{D}^0\rightarrow K^0_{S,L}\pi^+\pi^-$ and its impact on the measurement
  of the CKM angle $\gamma/\phi_3$}},
  }{}\href{https://doi.org/10.1103/PhysRevD.101.112002}{Phys.\ Rev.\ D
  \textbf{101} (2020) 112002},
  \href{http://arxiv.org/abs/2003.00091}{{\normalfont\ttfamily
  arXiv:2003.00091}}\relax
\mciteBstWouldAddEndPuncttrue
\mciteSetBstMidEndSepPunct{\mcitedefaultmidpunct}
{\mcitedefaultendpunct}{\mcitedefaultseppunct}\relax
\EndOfBibitem
\bibitem{Evans:2019wza}
T.~Evans, J.~Libby, S.~Malde, and G.~Wilkinson,
  \ifthenelse{\boolean{articletitles}}{\emph{{Improved sensitivity to the CKM
  phase $\gamma$ through binning phase space in $B^- \to DK^-$, $D \to
  K^+\pi^-\pi^-\pi^+$ decays}},
  }{}\href{https://doi.org/10.1016/j.physletb.2019.135188}{Phys.\ Lett.\ B
  \textbf{802} (2020) 135188},
  \href{http://arxiv.org/abs/1909.10196}{{\normalfont\ttfamily
  arXiv:1909.10196}}\relax
\mciteBstWouldAddEndPuncttrue
\mciteSetBstMidEndSepPunct{\mcitedefaultmidpunct}
{\mcitedefaultendpunct}{\mcitedefaultseppunct}\relax
\EndOfBibitem
\bibitem{BESIII:2021eud}
BESIII collaboration, M.~Ablikim {\em et~al.},
  \ifthenelse{\boolean{articletitles}}{\emph{{Measurement of the $D \to
  K^-\pi^+\pi^+\pi^-$ and $D \to K^-\pi^+\pi^0$ coherence factors and average
  strong-phase differences in quantum-correlated ${D\bar{D}}$ decays}},
  }{}\href{https://doi.org/10.1007/JHEP05(2021)164}{JHEP \textbf{05} (2021)
  164}, \href{http://arxiv.org/abs/2103.05988}{{\normalfont\ttfamily
  arXiv:2103.05988}}\relax
\mciteBstWouldAddEndPuncttrue
\mciteSetBstMidEndSepPunct{\mcitedefaultmidpunct}
{\mcitedefaultendpunct}{\mcitedefaultseppunct}\relax
\EndOfBibitem
\bibitem{Hsiao:2015cda}
Y.~K. Hsiao, P.~Y. Lin, C.~C. Lih, and C.~Q. Geng,
  \ifthenelse{\boolean{articletitles}}{\emph{{Charmful two-body anti-triplet
  $b$-baryon decays}},
  }{}\href{https://doi.org/10.1103/PhysRevD.92.114013}{Phys.\ Rev.\ D
  \textbf{92} (2015) 114013},
  \href{http://arxiv.org/abs/1509.05603}{{\normalfont\ttfamily
  arXiv:1509.05603}}\relax
\mciteBstWouldAddEndPuncttrue
\mciteSetBstMidEndSepPunct{\mcitedefaultmidpunct}
{\mcitedefaultendpunct}{\mcitedefaultseppunct}\relax
\EndOfBibitem
\bibitem{Wang:2008sm}
Y.-M. Wang, Y.~Li, and C.-D. Lu,
  \ifthenelse{\boolean{articletitles}}{\emph{{Rare Decays of $\Lambda_b \to
  \Lambda + \gamma$ and $\Lambda_b \to \Lambda + l^+ l^-$ in the Light-cone Sum
  Rules}}, }{}\href{https://doi.org/10.1140/epjc/s10052-008-0846-5}{Eur.\
  Phys.\ J.\ C \textbf{59} (2009) 861},
  \href{http://arxiv.org/abs/0804.0648}{{\normalfont\ttfamily
  arXiv:0804.0648}}\relax
\mciteBstWouldAddEndPuncttrue
\mciteSetBstMidEndSepPunct{\mcitedefaultmidpunct}
{\mcitedefaultendpunct}{\mcitedefaultseppunct}\relax
\EndOfBibitem
\bibitem{Wang:2015ndk}
Y.-M. Wang and Y.-L. Shen,
  \ifthenelse{\boolean{articletitles}}{\emph{{Perturbative Corrections to
  $\Lambda_b \to \Lambda$ Form Factors from QCD Light-Cone Sum Rules}},
  }{}\href{https://doi.org/10.1007/JHEP02(2016)179}{JHEP \textbf{02} (2016)
  179}, \href{http://arxiv.org/abs/1511.09036}{{\normalfont\ttfamily
  arXiv:1511.09036}}\relax
\mciteBstWouldAddEndPuncttrue
\mciteSetBstMidEndSepPunct{\mcitedefaultmidpunct}
{\mcitedefaultendpunct}{\mcitedefaultseppunct}\relax
\EndOfBibitem
\bibitem{LHCb-PAPER-2020-036}
LHCb collaboration, R.~Aaij {\em et~al.},
  \ifthenelse{\boolean{articletitles}}{\emph{{Measurement of CP observables in
  $B^\pm \to D^{(*)} K^{\pm}$ and $B^\pm \to D^{(*)} \pi^{\pm} $ decays using
  two-body $D$ final states}},
  }{}\href{https://doi.org/10.1007/JHEP04(2021)081}{JHEP \textbf{04} (2021)
  081}, \href{http://arxiv.org/abs/2012.09903}{{\normalfont\ttfamily
  arXiv:2012.09903}}\relax
\mciteBstWouldAddEndPuncttrue
\mciteSetBstMidEndSepPunct{\mcitedefaultmidpunct}
{\mcitedefaultendpunct}{\mcitedefaultseppunct}\relax
\EndOfBibitem
\bibitem{LHCb-PAPER-2020-019}
LHCb collaboration, R.~Aaij {\em et~al.},
  \ifthenelse{\boolean{articletitles}}{\emph{{Measurement of the CKM angle
  $\gamma$ in $B^{\pm} \to D K^{\pm}$ and $B^{\pm} \to D \pi^{\pm}$ decays with
  $D \to K_{\rm S} h^+h^-$}},
  }{}\href{https://doi.org/10.1007/JHEP02(2021)169}{JHEP \textbf{02} (2021)
  0169}, \href{http://arxiv.org/abs/2010.08483}{{\normalfont\ttfamily
  arXiv:2010.08483}}\relax
\mciteBstWouldAddEndPuncttrue
\mciteSetBstMidEndSepPunct{\mcitedefaultmidpunct}
{\mcitedefaultendpunct}{\mcitedefaultseppunct}\relax
\EndOfBibitem
\bibitem{LHCb-PAPER-2016-003}
LHCb collaboration, R.~Aaij {\em et~al.},
  \ifthenelse{\boolean{articletitles}}{\emph{{Measurement of \CP observables in
  \mbox{\decay{\Bpm}{\D \Kpm}} and \mbox{\decay{\Bpm}{\D\pipm}} with two- and
  four-body \D decays}},
  }{}\href{https://doi.org/10.1016/j.physletb.2016.06.022}{Phys.\ Lett.\
  \textbf{B760} (2016) 117},
  \href{http://arxiv.org/abs/1603.08993}{{\normalfont\ttfamily
  arXiv:1603.08993}}\relax
\mciteBstWouldAddEndPuncttrue
\mciteSetBstMidEndSepPunct{\mcitedefaultmidpunct}
{\mcitedefaultendpunct}{\mcitedefaultseppunct}\relax
\EndOfBibitem
\bibitem{LHCb-TDR-012}
LHCb collaboration, \ifthenelse{\boolean{articletitles}}{\emph{{Framework TDR
  for the LHCb Upgrade: Technical Design Report}}, }{}
  \href{http://cdsweb.cern.ch/search?p=CERN-LHCC-2012-007&f=reportnumber&action_search=Search&c=LHCb}
  {CERN-LHCC-2012-007}, 2012\relax
\mciteBstWouldAddEndPuncttrue
\mciteSetBstMidEndSepPunct{\mcitedefaultmidpunct}
{\mcitedefaultendpunct}{\mcitedefaultseppunct}\relax
\EndOfBibitem
\end{mcitethebibliography}


\end{document}